\documentclass[fleqn]{article}
\usepackage{graphicx}

\begin{document}

\title{
{\sf H\"older Inequalities and Bounds on the Masses of Light Quarks} }

\author{
T.G. Steele, K. Kostuik, J. Kwan\\
{\sl Department of Physics and Engineering Physics and}\\
{\sl Saskatchewan Accelerator Laboratory}\\
{\sl University of Saskatchewan}\\
{\sl Saskatoon, Saskatchewan S7N 5C6, Canada.}
}
\maketitle
\begin{abstract}
QCD Laplace sum-rules must satisfy a fundamental (H\"older) inequality
if they are to consistently represent an integrated hadronic cross-section.
After subtraction of the pion-pole, the Laplace sum-rule of pion currents
is shown to violate this fundamental inequality unless the $u$ and $d$ quark masses
(respectively denoted by $m_u$ and $m_d$)
are sufficiently large, placing a lower bound on the
$1 {\rm GeV}$ $\overline{\rm MS}$
running masses.
\end{abstract}
QCD Laplace sum-rules, a technique 
 which  equates a theoretical quantity to an integrated hadronic cross-section
to determine a QCD prediction of hadronic properties \cite{SVZ},
 have demonstrated their utility in numerous 
applications to hadronic physics.  It has recently been shown that 
Laplace sum-rules must satisfy a fundamental inequality in order to consistently represent
an integrated hadronic cross-section \cite{TGS}. When applied to
the Laplace sum-rule related to the pion, a strong dependence on  the light quark masses 
$m_u$ and $m_d$ 
occurs in the analysis of this inequality because the pseudoscalar correlation is proportional to 
$\left(m_u+m_d\right)^2$. In this paper we employ this fundamental H\"older inequality to place
lower bounds on the sum of the  $u$ and $d\,$ $1.0\, {\rm GeV}$  $\overline{\rm MS}$
running masses.

Consider the correlation function of  pseudoscalar
currents with the quantum numbers of the pion:
\begin{eqnarray}
& &J_5(x)=\frac{1}{\sqrt{2}}\left(m_u+m_d\right)\left[\bar u(x)i\gamma_5u(x)-
\bar d(x)i\gamma_5d(x)\right]
\label{j5}\\
& & \Pi_5(Q^2)=i\int d^4x\,e^{i q\cdot x}\langle O\vert T J_5(x) J_5(0)\vert O\rangle
\label{corrfn}
\end{eqnarray}
To leading order in the quark masses, the perturbative contributions $\Pi_5^{pert}$ are known to four-loop order 
in the  $\overline{\rm MS}$ scheme
\cite{chetyrkin}.
\begin{eqnarray}
\Pi_5^{pert}(Q^2)&=&\frac{3Q^2}{8\pi^2}\left(m_u+m_d\right)^2\log{\left(\frac{Q^2}{\mu^2}\right)}
\nonumber\\
& &\left[
1+\frac{\alpha}{\pi}\left( a_0+a_1L\right)
+\left(\frac{\alpha}{\pi}\right)^2\left[
b_0+b_1L+b_2L^2
\right]
+\left(\frac{\alpha}{\pi}\right)^3\left[
c_0+c_1L+c_2L^2+c_3 L^3
\right]
\right]\label{pert}
\\
& &L\equiv \log\left(\frac{Q^2}{\mu^2}\right)
\quad ,\quad
a_0=\frac{17}{3}\quad ,\quad a_1=-1
\\
& &b_0=45.846\quad ,\quad b_1=-\frac{95}{6}\quad ,\quad b_2=\frac{17}{12}
\\
& &c_0=465.8463\quad ,\quad c_1=-194.2393 \quad ,\quad c_2=38.1667 \quad ,\quad c_3=-2.3020825
\end{eqnarray}
Divergent polynomials in $Q^2$ are ignored in (\ref{pert}) since they correspond to subtraction constants in 
dispersion relations which do  not contribute to  sum-rules.

In the QCD sum-rule approach, nonperturbative effects are parametrized by the QCD condensates representing 
infinite correlation-length vacuum effects \cite{SVZ}.  In addition to the QCD condensate contributions,
scalar and pseudoscalar correlation functions must also take into account the effects of instantons which represent 
vacuum effects with a finite correlation length \cite{EVS}. 
 Thus the nonperturbative contributions 
to $\Pi_5$ consist of the QCD condensate effects $\Pi_5^{cond}$ and instanton effects $\Pi_5^{inst}$.  

To leading order in the quark mass, $\Pi_5^{cond}$ is given by \cite{BNRY,BLP}
\begin{eqnarray}
\Pi_5^{cond}(Q^2)=\left(m_u+m_d\right)^2\left[\frac{\langle \alpha G^2\rangle }{8\pi Q^2}
-\frac{\langle m\bar q q\rangle}{Q^2}+\frac{\pi \langle {\cal O}_6\rangle}{4Q^4}\right]
\label{cond}
\end{eqnarray}   
where we have used $SU(2)$ symmetry for the dimension-four quark condensates ({\it i.e.} 
$(m_u+m_d)\langle \bar u u\rangle+ \bar d d\rangle)\equiv 4 m\langle \bar q q\rangle$),
and   $\langle {\cal O}_6\rangle$ denotes the dimension six quark condensates 
\begin{eqnarray}
\langle{\cal O}_6\rangle&\equiv& \alpha_s
\biggl[ 
\left(2\langle \bar u \sigma_{\mu\nu}\gamma_5
T^au\bar u \sigma^{\mu\nu}\gamma_5T^a u\rangle
+ u\rightarrow d\right)
 -4\langle \bar u \sigma_{\mu\nu}\gamma_5T^au\bar d 
 \sigma^{\mu\nu}\gamma_5 T^a d\rangle
\biggr. \nonumber\\
& &\qquad\qquad 
\biggl.+\frac{2}{3}
\langle \left(   
\bar u \gamma_\mu T^a u+\bar d \gamma_\mu T^a d 
\right)
\sum_{u,d,s}\bar q \gamma^\mu T^aq
\rangle\biggr]
\label{o6}
\end{eqnarray}
The vacuum saturation hypothesis \cite{SVZ} will be used as a reference value for $\langle{\cal O}_6\rangle$
\begin{equation}
\langle{\cal O}_6\rangle=f_{vs}\frac{448}{27}\alpha
\langle \bar q q\bar q q\rangle
=f_{vs}3\times 10^{-3} {\rm GeV}^6 
\label{o61}
\end{equation}
where $f_{vs}=1$ for exact vacuum saturation.  Larger values 
of effective dimension-six
operators found in \cite{dimsix1,dimsix2} imply that $f_{vs}$  could be as 
large as $f_{vs}=2$.
The quark condensate is determined by the GMOR relation \cite{GMOR}
\begin{equation}
\left(m_u+m_d\right)\langle \bar u u+\bar d d\rangle=4m\langle \bar q q\rangle=
-2f_\pi^2m_\pi^2
\end{equation}
where $f_\pi=93\,{\rm MeV}$.  We will use recent determinations
of the gluon condensate $\langle \alpha G^2\rangle$
\cite{narison2}
\begin{equation}
\langle \alpha G^2\rangle=\left(0.07\pm 0.01\right)\,{\rm GeV^4}
\label{aGG}
\end{equation}
However, it should be noted that there is some discrepancy between 
\cite{narison2} 
 and the smaller value $\langle\alpha G^2\rangle=\left( 0.047\pm 0.014\right)\,{\rm GeV^4}$ found in \cite{dimsix2}.

The direct single-instanton contribution $\Pi_5^{inst}$ in the instanton liquid model is \cite{EVS}
\begin{equation}
\Pi_5^{inst}(Q^2)=\frac{3}{2\pi^2}\left(m_u+m_d\right)^2Q^2
\left[K_{-1}\left(\rho_c\sqrt{Q^2}\right)\right]^2
\label{instanton}
\end{equation}
where $K_n(x)$ denotes the modified Bessel function and $\rho_c\approx 1/(600\,{\rm MeV})$ is a fundamental scale
in the instanton liquid model \cite{EVS}.

The correlation function $\Pi_5(Q^2)$ satisfies a twice-subtracted dispersion relation
\begin{eqnarray}
& &\Pi_5(Q^2)=\Pi_5(0)-Q^2{\Pi_5}'(0)+\frac{Q^4}{\pi}\int\limits_0^\infty \frac{Im\,\Pi_5(t)}{t^2(t+Q^2)}dt
\label{dispersion} 
\\
& &\Pi_5(Q^2)=\Pi_5^{pert}(Q^2)+\Pi_5^{cond}(Q^2)+\Pi_5^{inst}(Q^2)
\label{total_pi}
\end{eqnarray}
Laplace sum-rules are formed from the dispersion relation by applying the Borel transform operator $\hat B$ 
to  (\ref{dispersion})
\begin{equation}
\hat B=\lim_{\stackrel{N \rightarrow \infty~,~Q^2\rightarrow \infty}{Q^2/N\equiv M^2}}
\frac{1}{\Gamma(N)}\left(-Q^2\right)^N\left(\frac{d}{dQ^2}\right)^N
\label{borel}
\end{equation}
 This results in  the Laplace sum-rule which  exponentially suppresses  the high-energy region.
\begin{eqnarray}
{\cal R}_0(M^2)&\equiv& M^2\hat B\left[\Pi_5\left(Q^2\right)\right]
\label{bor_lap}
\\
{\cal R}_0(M^2)&=&\frac{1}{\pi}\int\limits_0^\infty Im\Pi_5(t) e^{-t/M^2}\,dt
\label{basicsr}
\end{eqnarray}
Using the expressions (\ref{pert}, \ref{cond}, \ref{instanton}, \ref{total_pi}) for the correlation function,
the definition of $\hat B$ and the results of \cite{DEKS}, the sum-rule ${\cal R}_0(M^2)$ is obtained
\begin{eqnarray}
{\cal R}_0(M^2)&=&\frac{3(m_u+m_d^2)^2M^4}{8\pi^2}\left(   
1+4.821098 \frac{\alpha}{\pi}+21.97646\left(\frac{\alpha}{\pi}\right)^2+53.14179\left(\frac{\alpha}{\pi}\right)^3
\right)
\nonumber\\
& &+\left(m_u+m_d\right)^2\left(
-\langle m\bar q q\rangle 
+\frac{1}{8\pi}\langle \alpha G^2\rangle
+\frac{\pi\langle{\cal O}_6\rangle}{4M^2}
\right)
\nonumber\\
& &+\left(m_u+m_d\right)^2
{3\rho_c^2 M^6\over{8 \pi^2}} e^{-\rho_c^2M^2/2 }
\left[   
  K_0\left( {\rho_c^2M^2/2} \right) +
       K_1\left( {\rho_c^2M^2/2} \right)
\right]
\label{sr}
\end{eqnarray}
Renormalization group improvement of (\ref{sr}) implies that $\alpha$, $m_u$ and $m_d$ are
 running quantities evaluated at the mass scale $M$ in the $\overline{{\rm MS}}$ scheme \cite{NR}.   

The running coupling can be directly related to the experimental value \cite{PDG}
\begin{equation}
\alpha\left(M_Z\right)=0.119\pm 0.002
\label{alpha_MZ}
\end{equation}
by using (\ref{alpha_MZ}) as an initial condition for the evolution of $\alpha$
to the scale $M$ via the 4-loop $\overline{{MS}}$  beta function
\cite{beta}
\begin{eqnarray}
& &\mu^2\frac{d a}{d \mu^2}=
\beta\left(\frac{\alpha}{\pi}\right)=
-a^2\sum_{i=0}^\infty \beta_i \left(\frac{\alpha}{\pi}\right)^i\quad ,\quad a\equiv\frac{\alpha}{\pi}
\label{evolution}
\\
& &\beta_0=\frac{11-\frac{2}{3}n_f}{4}
\quad ,\quad
\beta_1=\frac{102-\frac{38}{3}n_f}{16}\quad ,\quad
\beta_2=\frac{\frac{2857}{2}-\frac{5033}{18}n_f+\frac{325}{54}n_f^2}{64}
\\[5pt]
& &
\beta_3=114.23033-27.133944 n_f+1.5823791 n_f^2-5.8566958\times 10^{-3}n_f^3
\end{eqnarray}
The only subtlety in this approach is the location of flavour thresholds
where the number of effective flavour degrees of freedom $n_f$ change.  In general, matching conditions
must be imposed at these thresholds to relate QCD with   $n_f$ quarks to an effective theory
with $n_f-1$ light quarks and a decoupled heavy quark \cite{rodrigo}.  Using the matching threshold $\mu_{th}$ defined by
$m_q(\mu_{th})=\mu_{th}$, where $m_q$ is the running quark mass, the matching condition to three-loop order is 
\cite{chetyrkin2}
\begin{equation}
a^{(n_f-1)}\left(\mu_{th}\right)=
a^{(n_f)}\left(\mu_{th}\right)\left[1+0.1528 \left[a^{(n_f)}\left(\mu_{th}\right)\right]^2
+\left\{0.9721-0.0847\left(n_f-1\right)\right\}\left[a^{(n_f)}\left(\mu_{th}\right)\right]^3\right]
\label{matching}
\end{equation}
For energies above the charm threshold, $\alpha(M)$ depends only on the initial condition
(\ref{alpha_MZ}) and the bottom threshold.  The currently 
accepted range for the bottom threshold \cite{PDG}
\begin{equation}
4.1\, {\rm GeV}\le m_b(m_b)\le 4.4\,{\rm GeV}
\label{mb_pdg}
\end{equation}
has a negligible effect on the evolution in $\alpha(M)$ in comparison with the uncertainty  (\ref{alpha_MZ}) in
$\alpha\left(M_Z\right)$.

Similarly, the running quark masses can be parametrized by  their (unknown) value at $1\,{\rm GeV}$.  Defining
\begin{eqnarray}
& &m=\frac{1}{2}\left(m_u+m_d\right)\\
& & w(M)=\frac{m(M)}{m\left(1\,{\rm GeV}\right)}\quad ,\quad w(1\,{\rm GeV})=1
\end{eqnarray}
allows the quark mass at the scale $M$ to be determined via the four-loop $\overline{{\rm MS}}$  scheme
anomalous mass dimension \cite{gamma}
\begin{eqnarray}
& &\frac{dw}{d\mu}=w(\mu)\frac{\gamma\left[ \frac{\alpha(\mu)}{\pi} \right]}{\beta\left[ \frac{\alpha(\mu)}{\pi} \right] }
\\
& &\gamma(x) = -x[1 + \sum_{i=1} \gamma_i x^i]
\\
& &\gamma_1 = 4.20833 - 0.138889n_f
\quad ,\quad
\gamma_2 = 19.5156 - 2.28412n_f - 0.0270062n_f^2
\\
& &\gamma_3 = 98.9434 - 19.1075 n_f + 0.276163 n_f^2 - 0.00579322 n_f^3
\end{eqnarray}
To evaluate the quark mass above the charm threshold the currently accepted values in the $\overline{\rm MS}$ scheme 
\cite{PDG} will be used
\begin{equation}
1.1\,{\rm GeV}\le m_c(m_c)\le 1.4\,{\rm GeV}
\label{m_c}
\end{equation}

In sum-rule applications,  
the $M$  dependence of the theoretical result ${\cal R}_0(M^2)$ (\ref{sr})
over a range of $M$ values is used to predict the resonance parameters
contained in
 $Im\Pi(t)$
through the relation (\ref{basicsr}).  
Since $Im\Pi(t)$ is in general related to a physical cross-section, 
fundamental inequalities for Laplace sum-rules can be constructed
using the property that $Im\Pi(t)>0$.  
Thus the simplest inequality that can be obtained is
\begin{equation}
{\cal R}_0(M^2)=\int Im\,\Pi(t)\,e^{-t/M^2}\,dt\ge 0
\longrightarrow \,{\cal R}_0(M^2)\ge 0
\end{equation}
More stringent sum-rule inequalities can be developed using 
integral inequalities.  In particular, H\"older inequalities for Laplace sum-rules
have been shown to be valuable in studying self-consistency of QCD sum-rules
and in obtaining bounds on the electromagnetic polarizability of charged pions 
\cite{TGS}.  H\"older's inequality for integrals defined over a measure 
$d\mu$ is \cite{holder}
\begin{eqnarray}
\biggl|\int_{t_1}^{t_2} f(t)g(t) d\mu \biggr| &\le& 
\left(\int_{t_1}^{t_2} \big|f(t)\big|^ p d\mu \right)^{1/p}
\left(\int_{t_1}^{t_2} \big|g(t)\big|^q d\mu \right)^{1/q}, \nonumber \\
&& \label{holder} \\
\frac{1}{p}+\frac{1}{q} &=&1~;\quad p,~q\ge 1\quad . \nonumber
\end{eqnarray}
When $p=q=2$ the H\"older inequality reduces to the well known Schwarz 
inequality. The key idea in applying H\"older's inequality to sum-rules is
recognizing that since
$Im\, \Pi(t)$ is positive 
it can serve as the measure
$d\mu=Im\,\Pi(t) dt$ in (\ref{holder}).
With the definitions
\begin{eqnarray}
& &\tau=\frac{1}{M^2}
\\
& &{\cal S}_k(\tau)\equiv\int\limits_{\mu_{th}}^\infty
Im\Pi(t) t^ke^{-t\tau} \,dt
\end{eqnarray}
(\ref{holder}) with $d\mu=Im\Pi(t) dt$, $f(t)=t^\alpha
e^{-at\tau}$, $g(t)=t^\beta e^{-bt\tau}$, leads to  
 the following inequalities  for ${\cal S}_k(\tau)$
\begin{equation}
{\cal S}_{\alpha+\beta}(\tau)\le
{\cal S}^{1/p}_{\alpha p}(ap\tau)
{\cal S}^{1/q}_{\beta q}(bq\tau)
\;;\quad a+b=1\quad .
\label{rat1}
\end{equation}
Imposing restrictions that we have the integer value $k=0$ 
in our sum-rule
(\ref{basicsr}) leads to the following inequality.
\begin{eqnarray}
{\cal S}_0[\omega\tau_{min}+(1-w)\tau_{max},s_0] &\le&
{\cal S}^\omega_0[\tau_{min},s_0] {\cal S}^{1-\omega}_0[\tau_{max},s_0],
\label{ineqa} \\
0\le\omega\le 1~&;&\quad 
\delta\tau=\tau_{max}-\tau_{min}>0
\end{eqnarray}

If the pion pole contribution to $Im\Pi_5(t)$ is explicitly included, then
the sum-rule (\ref{basicsr}) becomes
\begin{equation}
{\cal R}_0(\tau)=2f_\pi^2m_\pi^4e^{-m_\pi^2\tau}+\int\limits_{m_\pi^2}^\infty Im\Pi_5(t) e^{-t\tau}\, dt
\end{equation}
where $m_\pi$ represents the pion mass and $f_\pi=93\,{\rm MeV}$
is the pion decay constant.
Since for any reasonable range of $\tau$, $\exp{\left(-m_\pi^2\tau\right)}\approx 1$,
we find
\begin{equation}
{\cal S}_0(\tau)\equiv {\cal R}_0(\tau)-2f_\pi^2m_\pi^4=
\int\limits_{m_\pi^2}^\infty Im\Pi_5(t) e^{-t\tau}\, dt
\label{SSR}
\end{equation}
Thus if the sum-rule is a valid and consistent representation of 
the integration of $Im\Pi(t)$ in (\ref{basicsr}) then after subtraction of the pion pole, the sum-rule 
${\cal S}_0(\tau)$ must satisfy the  fundamental inequality
(\ref{ineqa}).
\begin{equation}
\rho_0 \equiv\frac{{\cal S}_0[\tau+(1-
\omega)\delta\tau]}{{\cal S}_0^\omega[\tau]
{\cal S}_0^{1-\omega}[\tau+\delta\tau]} \le 1
\quad ,\quad  \forall ~0\le \omega \le 1
\label{rat2a}
\end{equation}
Provided that $\delta\tau$ is reasonably small (in QCD $\delta\tau\approx 0.1\,GeV^{-2}$
appears sufficient)
these inequalities are  insensitive to the value of $\delta\tau$,
permitting a simple analysis of the inequality as a function of the energy-scale $M$.

The first applications of inequalities to quark-mass bounds used the
simple positivity constraint  ${\cal S}_0(\tau)\ge 0$ \cite{BNRY,NR,LRT}.  
We extend this analysis by using the more stringent H\"older inequality, higher-loop perturbative corrections, 
and inclusion of instanton effects  which are important for the 
pseudoscalar channel.

\begin{figure}
\centering
\includegraphics[scale=0.7]{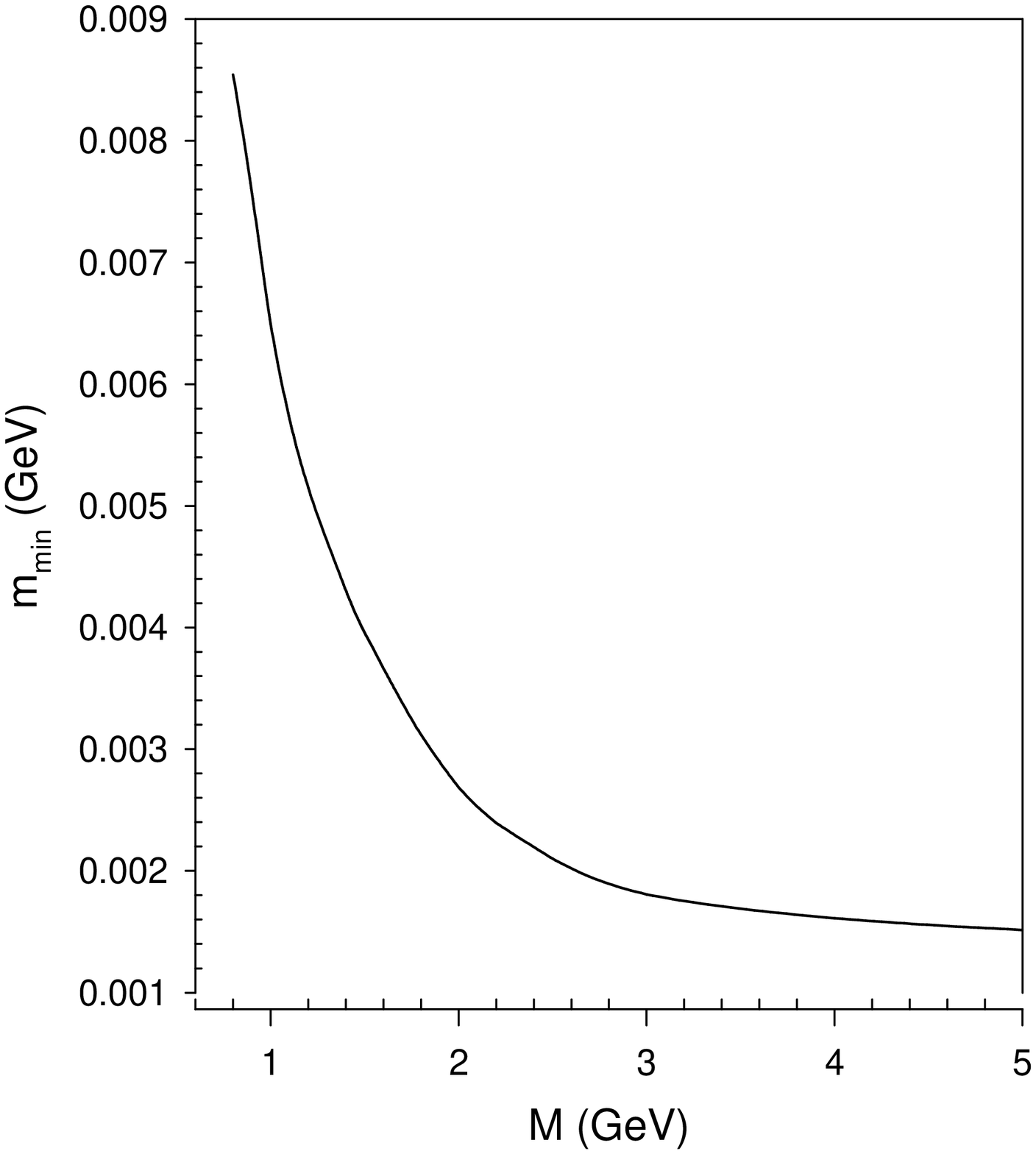}
\caption{
The quantity  $m_{min}$, representing the minimum
value of $m$ for which the
 H\"older inequality
(\protect\ref{rat2a}) is satisfied, is plotted as a function of the mass scale $M$
using central values of all input parameters.
}
\label{newfig1}
\end{figure}

\begin{figure}
\centering
\includegraphics[scale=0.7]{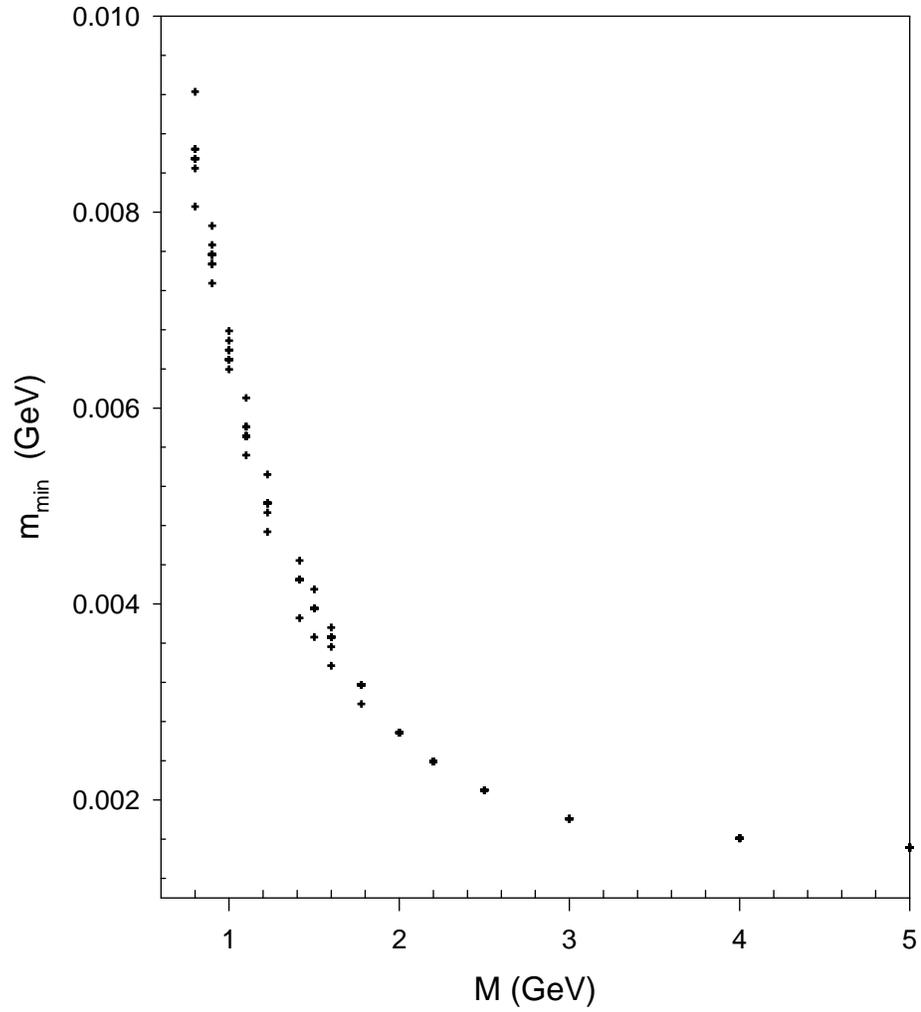}
\caption{
The effect of uncertainties in the input parameters on $m_{min}$
 is plotted as a function of the mass scale $M$.
Higher-loop effects are estimated by the inclusion of the five-loop perturbative
correction in (\protect\ref{sr2}).
}
\label{newfig2}
\end{figure}

For a fixed value of $M=1/\sqrt{\tau}$ the H\"older inequality can be used to find the
minimum value  of $m(1\,{\rm GeV})\equiv[m_u(1\,{\rm GeV})+m_d(1\,{\rm GeV}]/2 $ for which the 
inequality (\ref{rat2a}) is satisfied.  This lower bound on $m(1\,{\rm GeV})$ depends on the energy  scale
$M$ as indicated in Figure \ref{newfig1}.  However, 
the increasingly large coefficients in the perturbative portion of (\ref{sr}) 
suggests that these mass bounds could depend strongly upon higher-order perturbative effects.
Asymptotic Pad\'e approximation methods \cite{apap}  applied to the perturbative part of (\ref{sr}) result in the following estimate  
of ${\cal R}_0(M^2)$ with perturbative effects to five-loop order.
\begin{eqnarray}
{\cal R}_0(M^2)&=&\frac{3(m_u+m_d^2)^2M^4}{8\pi^2}\left(   
1+4.821098 \frac{\alpha}{\pi}+21.97646\left(\frac{\alpha}{\pi}\right)^2+53.14179\left(\frac{\alpha}{\pi}\right)^3
+137.6  \left(\frac{\alpha}{\pi}\right)^4 \right)
\nonumber\\
& &+\left(m_u+m_d\right)^2\left(
-\langle m\bar q q\rangle 
+\frac{1}{8\pi}\langle \alpha G^2\rangle
+\frac{\pi\langle{\cal O}_6\rangle}{4M^2}
\right)
\nonumber\\
& &+\left(m_u+m_d\right)^2
{3\rho_c^2 M^6\over{8 \pi^2}} e^{-\rho_c^2M^2/2 }
\left[   
  K_0\left( {\rho_c^2M^2/2} \right) +
       K_1\left( {\rho_c^2M^2/2} \right)
\right]
\label{sr2}
\end{eqnarray}
By comparing mass bounds with and without inclusion of this estimated five-loop perturbative  contribution, the
uncertainty in the mass bounds of Figure \ref{newfig1} devolving from truncation of the perturbative expansion can be estimated.
In addition to this perturbative uncertainty, 
a 50\% uncertainty in the vacuum saturation hypothesis (parameterized by $f_{vs}$), a 15\% uncertainty in the instanton size
$\rho_c$, and 
the effects of varying the input parameters within the ranges
(\ref{aGG},\ref{alpha_MZ},\ref{mb_pdg},\ref{m_c})  must also be considered.    
Figure \ref{newfig2} illustrates the effect of all sources of uncertainty upon the mass bounds obtained from the 
H\"older inequality. 
As indicated by the figure, the analysis is extremely stable.

The increase of the quark mass bound $m_{min}$ with decreasing $M$ implies that the minimum energy scale $M$ at 
which the pseudoscalar sum-rule is considered valid will 
provide the most stringent bound on the quark mass $m$.  
 The analysis of the $\tau$ hadronic width \cite{braaten}, and  hadronic contributions to $\alpha_{EM}(M_Z)$ and
the anomalous magnetic moment of the muon \cite{davier}
  provides excellent evidence for the the validity of QCD sum-rule methods at the energy scale $M_\tau\approx 1.8\,{\rm GeV}$. 
Furthermore,  the largest value of $M$ 
at which the uncertainties in the analysis are visible in Figure \ref{newfig2}  corresponds to $M\approx M_\tau$, 
providing support for the validity of the mass bounds at $M=M_\tau$.    
Thus a valid and {\em conservative} bound on the $1.0\,{\rm GeV}$ $\overline{{\rm MS}}$  
quark masses is obtained from Figure \ref{newfig2} at $M=M_\tau$
\begin{equation}
m(1\,{\rm GeV})=\frac{1}{2}\left[ m_u(1\,{\rm GeV})+m_d(1\,{\rm GeV})\right]\ge 3\,{\rm MeV}
\end{equation}
This conservative bound is phenomenologically significant since the Particle Data Group
quotes a lower bound of $m(1\, {\rm GeV})\ge 2.25\,{\rm MeV}$ \cite{PDG}.

\smallskip
\noindent
{\bf Acknowledgements:}  TGS is grateful for the financial support
of the Natural Sciences and Engineering Research Council of Canada (NSERC).

\clearpage

\clearpage

\end{document}